\documentclass[journal]{IEEEtran}

\usepackage{enumitem}
\usepackage[linesnumbered,ruled,lined]{algorithm2e}

\usepackage[
  colorlinks=true,
  linkcolor=blue,
  citecolor=blue,
  urlcolor=blue
]{hyperref}

\usepackage{booktabs,array,tabularx,makecell,threeparttable}
\newcolumntype{C}[1]{>{\centering\arraybackslash}m{#1}}
\newcolumntype{Y}{>{\centering\arraybackslash}X}

\AtBeginDocument{}

\usepackage{graphicx}
\usepackage{dcolumn}
\usepackage{bm}

\usepackage[usenames,dvipsnames]{xcolor}

\usepackage{amsmath,amsfonts}
\usepackage{mathtools} 

\usepackage{array}
\usepackage[caption=false,font=normalsize,labelfont=sf,textfont=sf]{subfig}
\usepackage{textcomp}
\usepackage{stfloats}
\usepackage{url}
\usepackage{verbatim}
\usepackage{cite}
\usepackage{svg}
\usepackage[noend]{algpseudocode} 
\usepackage{etoolbox}
\usepackage{multirow}

\usepackage{float}
\usepackage{pgfplots}
\usetikzlibrary{shapes.multipart,intersections}
\usepackage{amssymb,amsthm,steinmetz}
\usepackage{mathrsfs}  
\usepackage{acronym}
\usepackage{upgreek,xspace}

\usepackage{tikz}
\usetikzlibrary{calc}
\makeatletter
\newcommand{\gettikzxy}[3]{%
  \tikz@scan@one@point\pgfutil@firstofone#1\relax
  \edef#2{\the\pgf@x}%
  \edef#3{\the\pgf@y}%
}
\makeatother

\usepackage[draft]{todonotes}

\usepackage{esvect}

\usepackage{times}
\usepackage{stmaryrd}
\usepackage{babel}
\usepackage{graphics}
\usepackage{gensymb}

\ifCLASSINFOpdf

\else

\fi

\usepackage[all=normal,paragraphs=normal,floats=tight,mathspacing=normal,wordspacing=normal,charwidths=tight,mathdisplays=normal,leading=normal]{savetrees}

\begin{document}

\title{Mutual-Coupling-Aware Optimization of a Time-Floquet RIS for \\Harmonic Backscatter Communications}

\author{Aleksandr~D.~Kuznetsov,~\IEEEmembership{Graduate Student Member,~IEEE},~Ville~Viikari,~\IEEEmembership{Senior Member,~IEEE},~and~Philipp~del~Hougne,~\IEEEmembership{Member,~IEEE}
\thanks{
A.~D.~Kuznetsov, V.~Viikari, and P.~del~Hougne are with the Department of Electronics and Nanoengineering, Aalto University, 00076 Espoo, Finland. P.~del~Hougne is also with Univ Rennes, CNRS, IETR - UMR 6164, F-35000, Rennes, France. (e-mail: aleksandr.kuznetsov@aalto.fi; ville.viikari@aalto.fi; philipp.del-hougne@univ-rennes.fr)
\textit{(Corresponding Author: Philipp~del~Hougne.)}}
\thanks{This work was supported in part by Business Finland (projects RF ECO3 and MULTIRACS), the HPY Foundation (project 2026122), the Nokia Foundation (project 20260028), the ANR France 2030 program (project ANR-22-PEFT-0005), and the ANR PRCI program (project ANR-22-CE93-0010).}
}

\maketitle

\begin{abstract}
This Letter studies the optimization of a wireless communications system empowered by a periodically time-modulated reconfigurable intelligent surface, coined time-Floquet RIS (TF-RIS), in the presence of mutual coupling (MC) among the RIS elements. In contrast to a conventional RIS whose elements may be reconfigured between signaling intervals, a TF-RIS periodically modulates its elements within a signaling interval, thereby inducing frequency conversion. Periodic time modulation is particularly attractive for harmonic backscatter communications to avoid self-jamming. Based on time-Floquet multiport network theory, we formulate an MC-aware optimization problem for binary-amplitude-shift-keying (BASK) harmonic backscatter communications with practical 1-bit-programmable TF-RIS elements. We propose a general discrete-optimization algorithm and evaluate its performance based on realistic model parameters. We systematically examine the performance dependence on the time resolution of the periodic modulation and the number of retained harmonics. 
Benchmarking against an MC-unaware approach reveals the importance of MC awareness for the more challenging optimization problem of simultaneous desired-harmonic-channel-gain maximization and undesired-harmonic-channel-gain minimization.
\end{abstract}

\begin{IEEEkeywords}
Periodic time modulation, time-Floquet reconfigurable intelligent surface, mutual coupling, multiport network theory, harmonic beam and null steering, BASK, harmonic backscatter communications.
\end{IEEEkeywords}

\IEEEpeerreviewmaketitle

\section{Introduction}
\label{sec_introduction}

Backscatter communications date back at least to the 1940s~\cite{stockman1948communication,brooker2013lev} and are now ubiquitous in the form of RFID. Recent technological advances enable the fabrication of large arrays of backscatter elements, known as reconfigurable intelligent surfaces (RISs). Indeed, each RIS element can be understood as an antenna element whose port is terminated by a tunable load. 
Besides the common use of RISs for shaping wireless channels in conventional communications schemes, RISs also enable advanced forms of backscatter communications~\cite{zhao2020metasurface}. For instance, the large aperture and the large number of degrees of freedom of an RIS enable the precise steering of a backscatter signal to intended users, improving both information-transfer rate and security~\cite{zhao2020metasurface}. 

A fundamental challenge in backscatter communications relates to the fact that the received signal is the superposition of the information-carrying backscatter-modulated signal and other non-modulated signals. The latter can arise due to a direct line of sight between transmitter and receiver, due to multi-bounce paths that interact with environmental scattering objects but not the backscatter element(s), as well as due to the structural scattering of the backscatter element(s). In many practical scenarios, the non-modulated signal dominates, resulting in a form of self-jamming: a receiver with limited dynamic range struggles to decode the backscatter-modulated signal due to the strong non-modulated signal.

A promising remedy to this self-jamming problem is so-called \textit{harmonic backscattering}~\cite{ye2023review}. When the backscatter-modulated signal is shifted to a different carrier frequency than the transmitted signal, frequency orthogonality enables the receiver to capture the backscatter signal without any self-jamming. Common mechanisms for generating a backscatter-modulated signal at a different frequency involve the creation of harmonic signals based on non-linearity or periodic time-modulation of the backscatter element(s). We focus on periodic time modulation in this Letter. 
A conventional backscatter element is reconfigurable in the sense that its load state may be changed between signaling intervals, but it is fixed within a signaling interval. In contrast, a time-modulated backscatter element periodically modulates its load within a signaling interval, thereby enabling frequency conversion. Similarly, a conventional RIS may be reconfigured between signaling intervals, whereas a time-Floquet RIS (TF-RIS) periodically modulates its element states within a signaling interval.
As experimentally demonstrated in works such as~\cite{zhang2018space,xie2024thin}, a TF-RIS enables harmonic beam steering, which can enhance harmonic backscatter communications.

To the best of our knowledge, only a few theoretical works have studied wireless communications with TF-RISs to date~\cite{yurduseven2020intelligent,mizmizi2024wireless,verde2024rapidly}. These works use simplified models of TF-RISs that neglect important electromagnetic phenomena such as mutual coupling (MC). Meanwhile, a growing body of theoretical literature on conventional RISs is dedicated to MC-aware modeling and optimization~\cite{gradoni2021end,faqiri2022physfad,rabault2024tacit,matteo_universal}. Yet, we are not aware of any work on MC-aware optimization of TF-RISs.

In this Letter, we address this research gap. We consider for concreteness a scenario with a single transmitter, a single receiver, and a practical TF-RIS with 1-bit-programmable elements. Capitalizing on an MC-aware formalism for time-modulated systems based on time-Floquet multiport-network theory (TF-MNT), we study in a physics-consistent manner the optimization of the TF-RIS for binary-amplitude-shift-keying (BASK) harmonic backscatter communications. 

Our contributions are summarized as follows.
\textit{First}, we formulate the optimization problem for harmonic backscatter BASK with a TF-RIS in a physics-consistent manner based on TF-MNT. We also formulate a variant of this problem that simultaneously minimizes the harmonic signal at an undesired receiver.
\textit{Second}, we propose a general and efficient discrete optimization algorithm for 1-bit-programmable TF-RISs with MC.
\textit{Third}, we present a performance analysis based on a numerical full-wave simulation of a practical TF-RIS design. 
\textit{Fourth}, we benchmark our MC-aware approach against an MC-unaware one in terms of the achieved performance as a function of the number of retained harmonics and the time resolution of the periodic modulation.

\textit{Notation:}
$\mathbf{A} = \mathrm{diag}(\mathbf{a})$ denotes the diagonal matrix $\mathbf{A}$ whose diagonal entries are $\mathbf{a}$.
$\mathrm{blkdiag}(\mathbf{A}_1,\dots,\mathbf{A}_a)$ denotes the block-diagonal matrix constructed from the matrices $\mathbf{A}_1,\dots,\mathbf{A}_a$.
$\mathbf{A}_\mathcal{BC}$ denotes the block of the matrix $\mathbf{A}$ whose row [column] indices are in the set $\mathcal{B}$ [$\mathcal{C}$]. 
$|\mathcal{B}|$ denotes the cardinality of the set $\mathcal{B}$.
$^\top$ denotes the transpose operation.
$\mathbf{I}_a$ is the $a\times a$ identity matrix.
$\mathbf{0}$ denotes a matrix or vector of suitable size whose entries are all zero.
$\jmath$ is the imaginary unit.

\section{System Model}
\label{sec_system_model}

We briefly summarize the MNT model for a conventional RIS before generalizing it to TF-MNT for the TF-RIS.

\textit{Conventional RIS:} The RIS-parametrized radio environment is partitioned into $N_\mathrm{A}$ antenna ports via which waves are injected and/or received, $N_\mathrm{S}$ tunable lumped elements (typically, one per RIS element), and an ensemble of static scattering objects (including environmental scattering and structural scattering of antennas and RIS elements). Each tunable lumped element can be viewed as a ``virtual'' port terminated by a tunable load. Thus, the ensemble of static scattering objects is an $N$-port scattering system characterized by its scattering matrix $\mathbf{S}\in\mathbb{C}^{N\times N}$, where $N=N_\mathrm{A}+N_\mathrm{S}$. Furthermore, the ensemble of tunable individual loads constitutes an $N_\mathrm{S}$-port scattering system characterized by its scattering matrix $\mathbf{\Phi}\in\mathbb{C}^{N_\mathrm{S}\times N_\mathrm{S}}$, where $\mathbf{\Phi} = \mathrm{diag}(\mathbf{r})$, $\mathbf{r}=[r_1, r_2, \dots, r_{N_\mathrm{S}}]$, and $r_i\in\mathbb{C}$ is the reflection coefficient of the tunable load associated with the $i$th RIS element. We define all scattering parameters with a reference impedance of $Z_0=50\ \Omega$, and we assume that signal generators and detectors are matched to $Z_0$.
The $N_\mathrm{S}$ ``virtual'' ports of $\mathbf{S}$ are terminated by $\mathbf{\Phi}$. According to standard MNT, the end-to-end channel matrix $\mathbf{H}\in\mathbb{C}^{N_\mathrm{R}\times N_\mathrm{T}}$ from $N_\mathrm{T}$ transmitting antennas to $N_\mathrm{R}=N_\mathrm{A}-N_\mathrm{T}$ distinct receiving antennas is given by~\cite{matteo_universal} 
\begin{equation}
\mathbf{H}
= \mathbf{S}_{\mathcal{RT}}
+ \mathbf{S}_{\mathcal{RS}}\,
\bigl(\mathbf{I}_{N_\mathrm{S}}-\mathbf{\Phi}\,\mathbf{S}_{\mathcal{SS}}\bigr)^{-1}
\,\mathbf{\Phi}\,
\mathbf{S}_{\mathcal{ST}},
\label{eq1}
\end{equation}
where $\mathcal{T}$, $\mathcal{R}$, and $\mathcal{S}$ denote the sets of port indices associated with transmitting antennas, receiving antennas, and RIS elements. The mapping from an input wavefront $\mathbf{a}\in\mathbb{C}^{N_\mathrm{T}}$ to the corresponding output wavefront $\mathbf{b}\in\mathbb{C}^{N_\mathrm{R}}$ is $\mathbf{H}$:
\begin{equation}
    \mathbf{b} = \mathbf{H}\, \mathbf{a}.
    \label{eq_bHa}
\end{equation}
MC between RIS elements is captured by the off-diagonal entries of $\mathbf{S}_\mathcal{SS}$ and results in a strongly non-linear mapping from $\mathbf{r}$ to $\mathbf{H}$. Assuming $\mathbf{S}_\mathcal{SS}=\mathbf{0}$ thus implies the absence of MC and simplifies (\ref{eq1}) to the widespread cascaded model:
\begin{equation}
    \mathbf{H}^\mathrm{CASC} = {\mathbf{S}}_\mathcal{RT} +{\mathbf{S}}_\mathcal{RS}\, \mathbf{\Phi}\, {\mathbf{S}}_\mathcal{ST},
    \label{eqCASC}
\end{equation}
where the mapping from $\mathbf{r}$ to $\mathbf{H}$ is affine.

\begin{figure}
    \centering
    \includegraphics[width=\columnwidth]{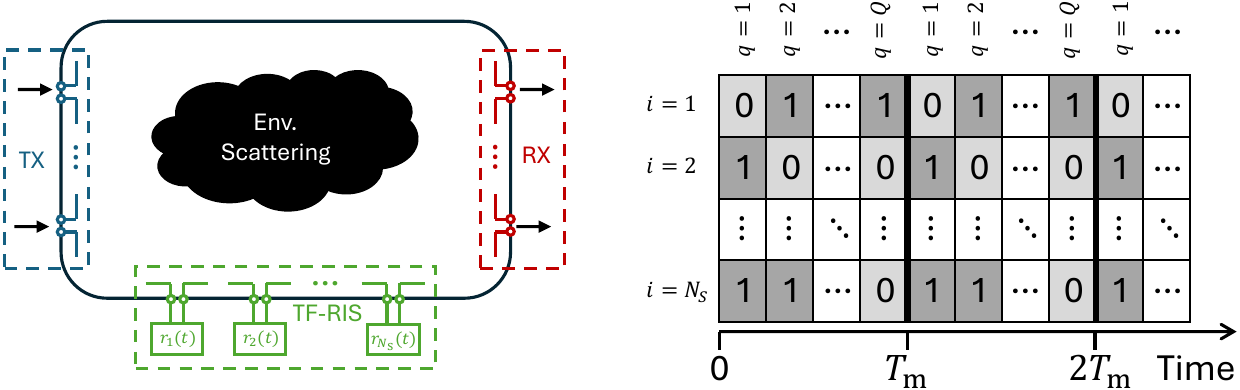}
    \caption{Left: MNT-based system model for a TF-RIS-parametrized radio environment. Right: Binary time-periodic load-modulation pattern.}
    \label{Fig1}
\end{figure}

\textit{TF-RIS:} Periodic time modulation results in frequency conversion. 
Analogous to before, we partition a TF-RIS-parametrized radio environment into antenna ports, tunable lumped elements, and static scattering objects. Frequency conversion can only occur at the time-modulated loads of the RIS elements, because all other parts of the system are linear and time-invariant. Assuming a periodic modulation of the loads with modulation frequency $f_\mathrm{m}$, a signal incident at $f_0 \gg f_\mathrm{m}$ can scatter into the harmonics $f_h = f_0 + h f_\mathrm{m}$, where $h$ is an integer. In practice, only a finite set of harmonics $\mathcal H=\{h_1,\dots,h_{|\mathcal H|}\}\subset\mathbb{Z}$ is retained. In this Letter, we always choose $\mathcal{H}$ symmetric around the fundamental harmonic $h=0$, such that $0\in\mathcal{H}$ and $|\mathcal{H}|$ is odd.
At each frequency $f_h$, scattering between the $N_\mathrm{A}$ antenna ports and the $N_\mathrm{S}$ ``virtual'' ports is linear and time-invariant, and hence characterized by $\mathbf{S}^{(h)}\in\mathbb{C}^{N\times N}$, where $^{(h)}$ indicates the considered harmonic. 

We now seek the end-to-end mapping from the transmitted signal at $f_0$ to the received harmonic signal at $f_h$, as a function of the loads' periodic time modulation. Following~\cite{kuznetsov2025multifrequency}, we begin by defining an augmented multi-frequency input wavefront $\widetilde{\mathbf{a}} = \left[{\mathbf{a}^{(h_1)}}^\top {\mathbf{a}^{(h_2)}}^\top \dots \ {{\mathbf{a}}^{(h_{|\mathcal H|})}}^\top \right]^\top \in \mathbb{C}^{|\mathcal H| N_\mathrm{T}}$, where $\mathbf{a}^{(h_i)}\in\mathbb{C}^{N_\mathrm{T}}$ is the input wavefront at frequency $f_{h_i}$. Analogously, we define the augmented multi-frequency output wavefront $\widetilde{\mathbf{b}}\in\mathbb{C}^{|\mathcal H| N_\mathrm{R}}$. 
For a fixed periodic load-modulation pattern, the TF-RIS-parametrized radio environment is a linear periodically time-varying system; in the augmented multi-frequency domain, it is therefore represented by a linear map. Hence, analogous to (\ref{eq_bHa}), the mapping from $\widetilde{\mathbf{a}}$ to $\widetilde{\mathbf{b}}$ is linear:
\begin{equation}
\widetilde{\mathbf{b}} = \widetilde{\mathbf{H}}\,\widetilde{\mathbf{a}},
\end{equation}
where $\widetilde{\mathbf{H}}\in\mathbb{C}^{|\mathcal H|N_\mathrm{R}\times |\mathcal H|N_\mathrm{T}}$ is the augmented multi-frequency end-to-end channel matrix.
For a conventional RIS without periodic time modulation, $\widetilde{\mathbf{H}}$ would be block-diagonal because different harmonics would not couple to each other. In contrast, for a TF-RIS, $\widetilde{\mathbf{H}}$ is generally \textit{not} block-diagonal.

Because the system excluding the loads is linear and time-invariant, it does not couple different harmonics. Therefore, the associated augmented scattering blocks are block-diagonal:
\begin{equation}
\widetilde{\mathbf S}_{\mathcal{XY}}
=
\mathrm{blkdiag}\!\left(
\mathbf S_{\mathcal{XY}}^{(h_1)},
\mathbf S_{\mathcal{XY}}^{(h_2)},
\dots,
\mathbf S_{\mathcal{XY}}^{(h_{|\mathcal H|})}
\right),
\end{equation}
where $\mathcal X,\mathcal Y\in\{\mathcal T,\mathcal R,\mathcal S\}$.
In contrast, the augmented load matrix
\begin{equation}
\widetilde{\mathbf\Phi}
=
\begin{bmatrix}
\mathbf\Phi^{(h_1,h_1)} & \cdots & \mathbf\Phi^{(h_1,h_{|\mathcal H|})}\\
\vdots & \ddots & \vdots\\
\mathbf\Phi^{(h_{|\mathcal H|},h_1)} & \cdots & \mathbf\Phi^{(h_{|\mathcal H|},h_{|\mathcal H|})}
\end{bmatrix}
\in\mathbb{C}^{|\mathcal H|N_\mathrm{S}\times |\mathcal H|N_\mathrm{S}}
\end{equation}
captures the harmonic coupling induced by periodic time modulation, where $\mathbf\Phi^{(h_n,h_m)}\in\mathbb{C}^{N_\mathrm{S}\times N_\mathrm{S}}$ maps waves impinging on the load network at $f_{h_m}$ to waves reflected by the load network at $f_{h_n}$. For independent element-wise modulation, each block $\mathbf\Phi^{(h_n,h_m)}$ is diagonal, with diagonal entries given by the corresponding Fourier coefficients of the time-varying load reflection coefficients, as discussed below.
Analogous to (\ref{eq1}), $\widetilde{\mathbf{H}}$  follows from standard MNT arguments:
\begin{equation}
\widetilde{\mathbf H}
=
\widetilde{\mathbf S}_{\mathcal{RT}}
+
\widetilde{\mathbf S}_{\mathcal{RS}}
\bigl(
\mathbf I_{|\mathcal H|N_\mathrm{S}}
-
\widetilde{\mathbf\Phi}\,
\widetilde{\mathbf S}_{\mathcal{SS}}
\bigr)^{-1}
\widetilde{\mathbf\Phi}\,
\widetilde{\mathbf S}_{\mathcal{ST}}.
\label{eq:Htilde}
\end{equation}
Analogous to (\ref{eqCASC}), assuming $\widetilde{\mathbf{S}}_\mathcal{SS}=\mathbf{0}$, which implies the absence of MC at all considered harmonics, a simplified cascaded system model for the TF-RIS follows:
\begin{equation}
\widetilde{\mathbf H}^\mathrm{CASC}
=
\widetilde{\mathbf S}_{\mathcal{RT}}
+
\widetilde{\mathbf S}_{\mathcal{RS}}\,
\widetilde{\mathbf\Phi}\,
\widetilde{\mathbf S}_{\mathcal{ST}}.
\label{eq:HtildeCASC}
\end{equation}

As mentioned, we consider 1-bit-programmable RIS elements, meaning that each load can realize only two distinct reflection states. To allow for dispersion, we denote by $\alpha^{(h)}\in\mathbb{C}$ and $\beta^{(h)}\in\mathbb{C}$ the frequency-dependent reflection coefficients of the two static load states at frequency $f_h$. As shown in Fig.~\ref{Fig1}, we divide one modulation period $T_\mathrm{m}=1/f_\mathrm{m}$ into $Q$ equal slots of duration $\Delta=T_\mathrm{m}/Q$. We assume that all RIS elements are synchronized to the same modulation clock, so that they share a common period $T_\mathrm{m}$ and the same $Q$ slot boundaries, while their binary states may differ from slot to slot and from RIS element to RIS element. The periodic switching pattern of the RIS is thus described by a binary configuration matrix $\mathbf{C}\in\{0,1\}^{N_\mathrm{S}\times Q}$, whose $(i,q)$th entry we denote by $C_{i,q}$. Within each slot, the load is frozen; the reflection coefficient of the $i$th load in slot $q$ at frequency $f_{h_m}$ is $r_i^{(q,h_m)}=\beta^{(h_m)}+(\alpha^{(h_m)}-\beta^{(h_m)})C_{i,q}$. Let $r_i^{(h_m)}(t)$ denote the corresponding periodic piecewise-constant reflection coefficient of the $i$th load over one modulation period at frequency $f_{h_m}$. The corresponding Fourier coefficient from incident harmonic $h_m$ to reflected harmonic $h_n$ is
\begin{equation}
\phi_i^{(h_n,h_m)}
=
\frac{1}{T_\mathrm{m}}
\sum_{q=1}^{Q}
\int_{(q-1)\Delta}^{q\Delta}
 r_i^{(q,h_m)}\mathrm{e}^{-\jmath 2\pi (h_n-h_m)t/T_\mathrm{m}}\,\mathrm{d}t,
\end{equation}
and, accordingly,
\begin{equation}
\mathbf\Phi^{(h_n,h_m)}
=
\operatorname{diag}\!\left(\left[
 \phi_1^{(h_n,h_m)},
 \dots,
 \phi_{N_\mathrm{S}}^{(h_n,h_m)}
 \right]\right).
\end{equation}
Altogether, given $\alpha^{(h)}$, $\beta^{(h)}$, and $\mathbf{S}^{(h)}$ for all $h\in\mathcal{H}$, as well as $T_\mathrm{m}$, our system model allows us to physics-consistently evaluate the end-to-end channel from any transmitter port at any harmonic to any receiver port at any harmonic, for any feasible $\mathbf{C}$.

\section{MC-Aware TF-RIS Optimization}

\textit{Problem Formulation:} For concreteness, we consider $N_\mathrm{T}=N_\mathrm{R}=1$. The transmitter radiates a unit-amplitude signal at $f_0$, i.e., $a^{(0)}=1$ and $a^{(h)}=0 \ \forall \ h\in\mathcal{H}\setminus\{0\}$. The received signal at the target harmonic $\hat{h} \neq 0$ is $b^{(\hat{h})}$. For the two BASK symbols, we seek two TF-RIS modulation patterns, $\mathbf{C}_\mathrm{+}$ and $\mathbf{C}_\mathrm{-}$, that maximize the received amplitude contrast at $f_{\hat h}$:
\begin{equation}
\tag{P1}
\max_{\mathbf{C}_{+},\,\mathbf{C}_{-}}
 \left|\, |b^{(\hat h)}(\mathbf{C}_{+})| - |b^{(\hat h)}(\mathbf{C}_{-})| \,\right|.
\label{eq:bask_opt}
\end{equation}

An optimal closed-form choice for $\mathbf{C}_\mathrm{-}$ is any pattern whose rows are constant over $q$, implying the absence of any time modulation and thus any frequency conversion, and ultimately $b^{(\hat{h})}(\mathbf{C}_\mathrm{-})=0$.
Hence, solving (\ref{eq:bask_opt}) simplifies to
\begin{equation}
\tag{P2}
\max_{\mathbf{C}_\mathrm{+}}
|b^{(\hat{h})}(\mathbf{C}_\mathrm{+})|.
\label{eq:on_only_opt}
\end{equation}

In addition, we consider the more challenging problem of simultaneous beam and null steering. Instead of just maximizing the harmonic signal at the intended receiver with $\mathbf{C}_+$, we simultaneously aim to minimize the harmonic signal at an unintended receiver. Possible motivations include interference mitigation and security. Nulling is known to be more challenging and sensitive to MC awareness~\cite{yven2025end}. In this scenario, we have $N_\mathrm{T}=1$, $N_\mathrm{R}=2$, and $N_\mathrm{A}=3$. We aim to maximize the difference between the harmonic signal amplitude $b^{(\hat{h})}$ at the intended receiver and the harmonic signal amplitude $\bar{b}^{(\hat{h})}$ at the unintended receiver:
\begin{equation}
\tag{P3}
\max_{\mathbf{C}_\mathrm{+}}
\left(|b^{(\hat{h})}(\mathbf{C}_\mathrm{+})|-|\bar{b}^{(\hat{h})}(\mathbf{C}_\mathrm{+})|\right).
\label{eq:on_only_opt_beamplusnull}
\end{equation}

\textit{MC-Aware Optimization Algorithm:}
The optimization problems in (\ref{eq:on_only_opt}) and (\ref{eq:on_only_opt_beamplusnull}) are discrete, non-convex, and high-dimensional combinatorial problems because the binary decision variables are the $N_\mathrm{S}Q$ entries of $\mathbf{C}_+$ and, under MC, the mapping from $\mathbf{C}_+$ to $b^{(\hat h)}$ (and $\bar{b}^{(\hat h)}$) is strongly non-linear through the matrix inverse in (\ref{eq:Htilde}). Exhaustive search over all $2^{N_\mathrm{S}Q}$ feasible modulation patterns is therefore intractable except for very small systems.\footnote{For amplitude-based objectives, patterns related by a common cyclic shift of the $Q$ time slots are equivalent. The number of distinct patterns is then $(1/Q)\sum_{\ell=0}^{Q-1}(2^{N_\mathrm{S}})^{\gcd(Q,\ell)}$, where $\gcd(Q,\ell)$ denotes the greatest common divisor of $Q$ and $\ell$. This count still grows exponentially with $Q N_\mathrm{S}$.}
For the MC-aware discrete optimization of conventional RISs, recent work found that coordinate descent offers an attractive trade-off between achieved performance, runtime, and memory~\cite{hammami2026statistical}. Motivated by this finding, we adopt a discrete steepest-coordinate-ascent strategy summarized in Algorithm~\ref{alg:cd}. 
At each iteration, all single-entry flips of the current configuration are evaluated, and the algorithm moves to the improving flip that yields the largest objective improvement. The procedure terminates when no single-entry flip improves the objective. $\mathcal{O}$ denotes the objective to be maximized. Specifically, we retain the best outcome across 100 randomly initialized runs of Algorithm~\ref{alg:cd} to avoid seed-dependent local optima.\footnote{The computational cost of Algorithm~\ref{alg:cd} can be reduced by updating, rather than recomputing, the inverse in (\ref{eq:Htilde}). Specifically, a flip of a single binary variable $C_{i,q}$ only changes the frequency-dependent periodic reflection coefficients $r_i^{(h)}(t)$ of the $i$th RIS element over the retained harmonics. Therefore, the induced perturbation in $\widetilde{\mathbf{\Phi}}$ is confined to the $|\mathcal H|$ augmented coordinates associated with the $i$th RIS element across the retained harmonics, and the resulting perturbation of $\mathbf{I}_{|\mathcal  H|N_\mathrm{S}}-\widetilde{\mathbf{\Phi}}\,\widetilde{\mathbf S}_{\mathcal{SS}}$ has rank at most $|\mathcal H|$. Consequently, a Woodbury-based inverse-update approach can replace for each candidate flip a full inversion of the $|\mathcal H|N_\mathrm{S}\times |\mathcal H|N_\mathrm{S}$ matrix, which dominates with a complexity of $\mathcal{O}((|\mathcal{H}|N_\mathrm{S})^3)$, by the inversion of a much smaller matrix of size at most $|\mathcal H|\times |\mathcal H|$ with complexity of $\mathcal{O}(|\mathcal{H}|^3)$~\cite{prod2023efficient}. This is particularly beneficial when $N_\mathrm{S}$ is large and $|\mathcal{H}|$ is moderate or small.}

\begin{algorithm}
\begingroup\footnotesize
\caption{{\small Binary Steepest-Coordinate Ascent}}
\label{alg:cd}
Initialize RIS configuration: \( \mathbf{C}_{\mathrm{hold}} \gets \mathbf{C}_{\mathrm{init}} \). \\
$\mathcal{O}_\mathrm{hold} \gets \mathcal{O}(\mathbf{C}_\mathrm{hold})$;
$f \gets 1$.\\
\While{\(f \neq 0 \)}{
    $\mathbf{C}_{\mathrm{curr}} \gets \mathbf{C}_{\mathrm{hold}}$;
    $k \gets 0$; $f \gets 0$.\\
    \While{\( k < N_{\mathrm{S}} \)}{
    $k \gets k+1$; $q \gets 0$.\\
    \While{\( q < Q \)}{
        $q \gets q  + 1$.\\
        \(\mathbf{C}_\mathrm{new} \gets \mathbf{C}_\mathrm{curr}\) with its $(k,q)$th entry flipped.\\
        $\mathcal{O}_\mathrm{new} \gets \mathcal{O}(\mathbf{C}_\mathrm{new})$.  \\
        \If{\( \mathcal{O}_\mathrm{new} > \mathcal{O}_\mathrm{hold} \)}{
         \( \mathbf{C}_{\mathrm{hold}} \gets \mathbf{C}_{\mathrm{new}} \); 
         \( \mathcal{O}_{\mathrm{hold}} \gets \mathcal{O}_{\mathrm{new}} \);
         \( f \gets 1 \).
        }
    }
    }
    
}
\textbf{Output:} Optimized RIS configuration \( \mathbf{C}_{\mathrm{curr}}. \)
\endgroup
\end{algorithm}

\section{Performance Evaluation}

\begin{figure}
    \centering
    \includegraphics[width=\columnwidth]{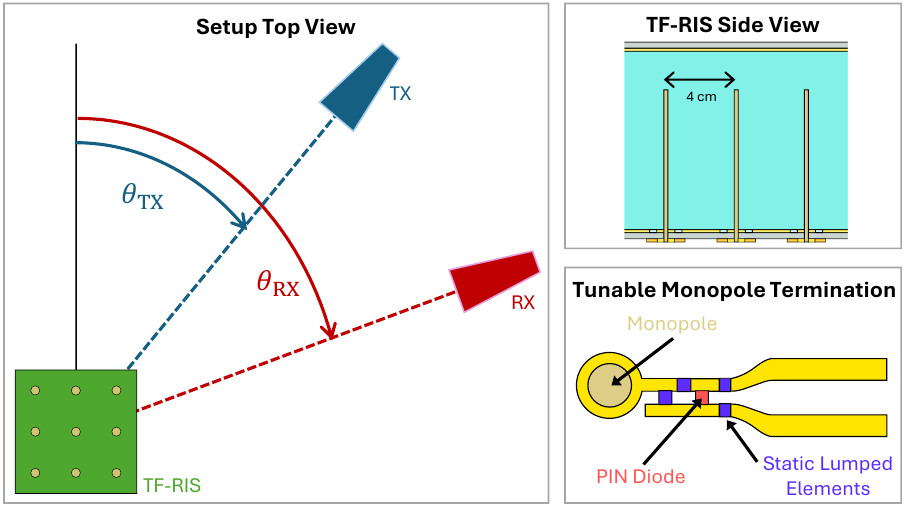}
    \caption{Considered TF-RIS setup based on~\cite{kuznetsov2025multifrequency}.}
    \label{Fig2}
\end{figure}

\textit{Setup:} Our considered quasi-2D setup is sketched in Fig.~\ref{Fig2} and is based on the practical TF-RIS design from~\cite{kuznetsov2025multifrequency}. Our small-scale TF-RIS consists of nine time-modulated RIS elements arranged in a regular $3\times3$ grid with one-third-wavelength spacing, and it simultaneously transmits and reflects. Each RIS element consists of a monopole terminated by a time-modulated PIN diode.   Further details on the RIS design can be found in~\cite{kuznetsov2025multifrequency}. We consider $f_0=2.4\ \mathrm{GHz}$, $f_\mathrm{m} = 100\ \mathrm{kHz}$, and $\hat{h}=1$ in this Letter. We extract $\alpha^{(h)}$ and $\beta^{(h)}$ from the datasheet of a commercial PIN diode (Skyworks SMP1345-079LF) and obtain $\mathbf{S}^{(h)}$ from full-wave numerical simulation (CST Studio Suite). 
We report channel gains with respect to plane-wave excitation/reception, excluding antenna directivity contributions.

\begin{figure}
    \centering
    \includegraphics[width=0.8\columnwidth]{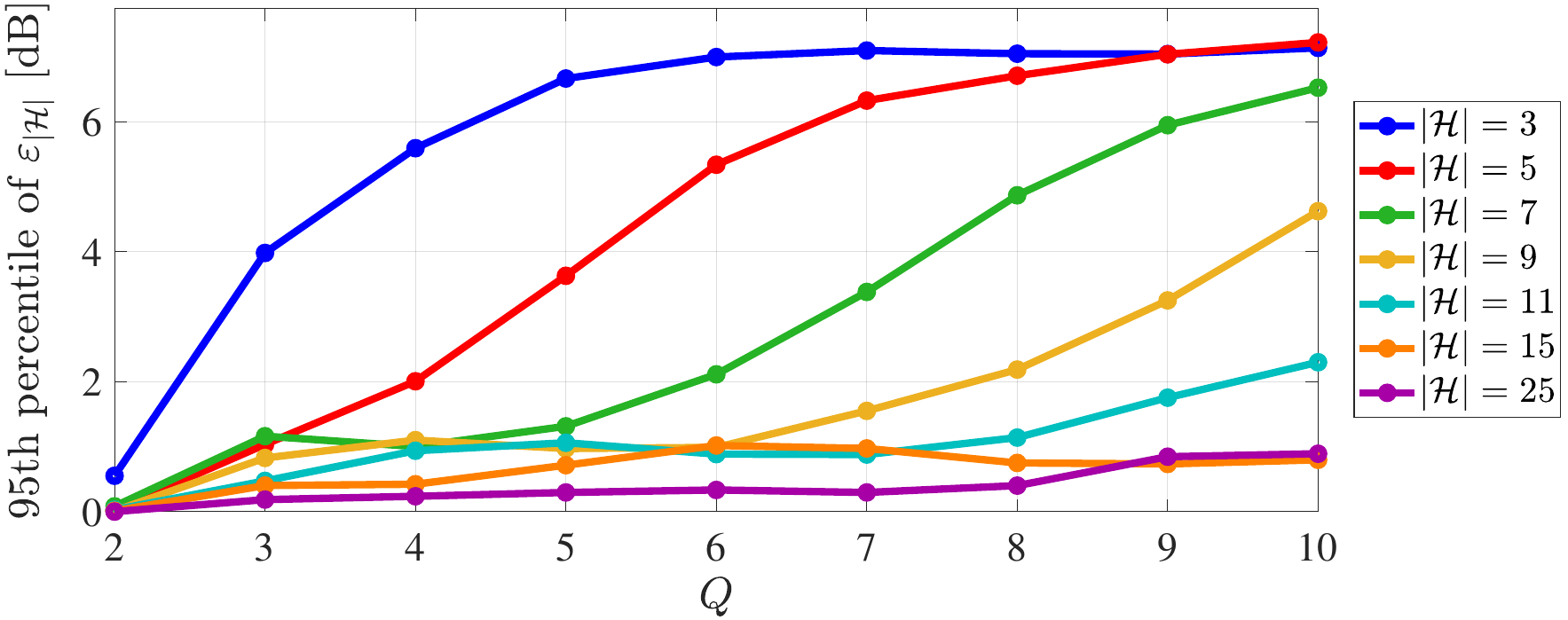}
    \caption{Effect of the number of retained harmonics $|\mathcal{H}|$ on the TF-MNT accuracy, as a function of $Q$. We quantify the TF-MNT accuracy in terms of the 95th percentile of $\varepsilon_{|\mathcal{H}|}$ (see main text for details).}
    \label{Fig3}
\end{figure}

\textit{Influence of Number of Retained Harmonics:}
In the MC-unaware case, omitted harmonics cannot influence the target harmonic through multiple harmonic-scattering events. In the MC-aware case, the retained harmonics participate in repeated inter-harmonic scattering through multiple interactions with the time-modulated loads. Thus, the harmonic truncation level $|\mathcal H|$ governs both the computational cost and the accuracy of the TF-MNT model. 
Let $b^{(\hat{h})}_{|\mathcal{H}|}$ denote the received signal at the target harmonic computed with $|\mathcal{H}|$ retained harmonics. We assess in Fig.~\ref{Fig3} how the model accuracy depends on $|\mathcal H|$. Our accuracy metric is the 95th percentile of $\varepsilon_{|\mathcal{H}|} = \left|20\log_{10}|b^{(\hat{h})}_{|\mathcal{H}|}| - 20\log_{10}|b^{(\hat{h})}_{51}|\right|$, evaluated for $\theta_\mathrm{TX}=40^\circ$, over all $\theta_\mathrm{RX}$ sampled in $10^\circ$ steps and over 10,000 random TF-RIS configurations. We take the case $|\mathcal H|=51$ as ground truth. We see in Fig.~\ref{Fig3} that larger values of $Q$ generally require larger $|\mathcal H|$ to achieve the same accuracy. For fixed $|\mathcal H|$, the 95th percentile of $\varepsilon_{|\mathcal{H}|}$ increases with $Q$ and eventually saturates around 7~dB. For $|\mathcal H|=3$ and $|\mathcal H|=5$, this percentile is already relatively large even for small $Q$. However, as seen below, this observation does not systematically preclude good optimization performance with small $|\mathcal H|$, because our accuracy metric emphasizes worst-case dB errors, which typically occur for low-gain channels.

\begin{figure}
    \centering
    \includegraphics[width=\columnwidth]{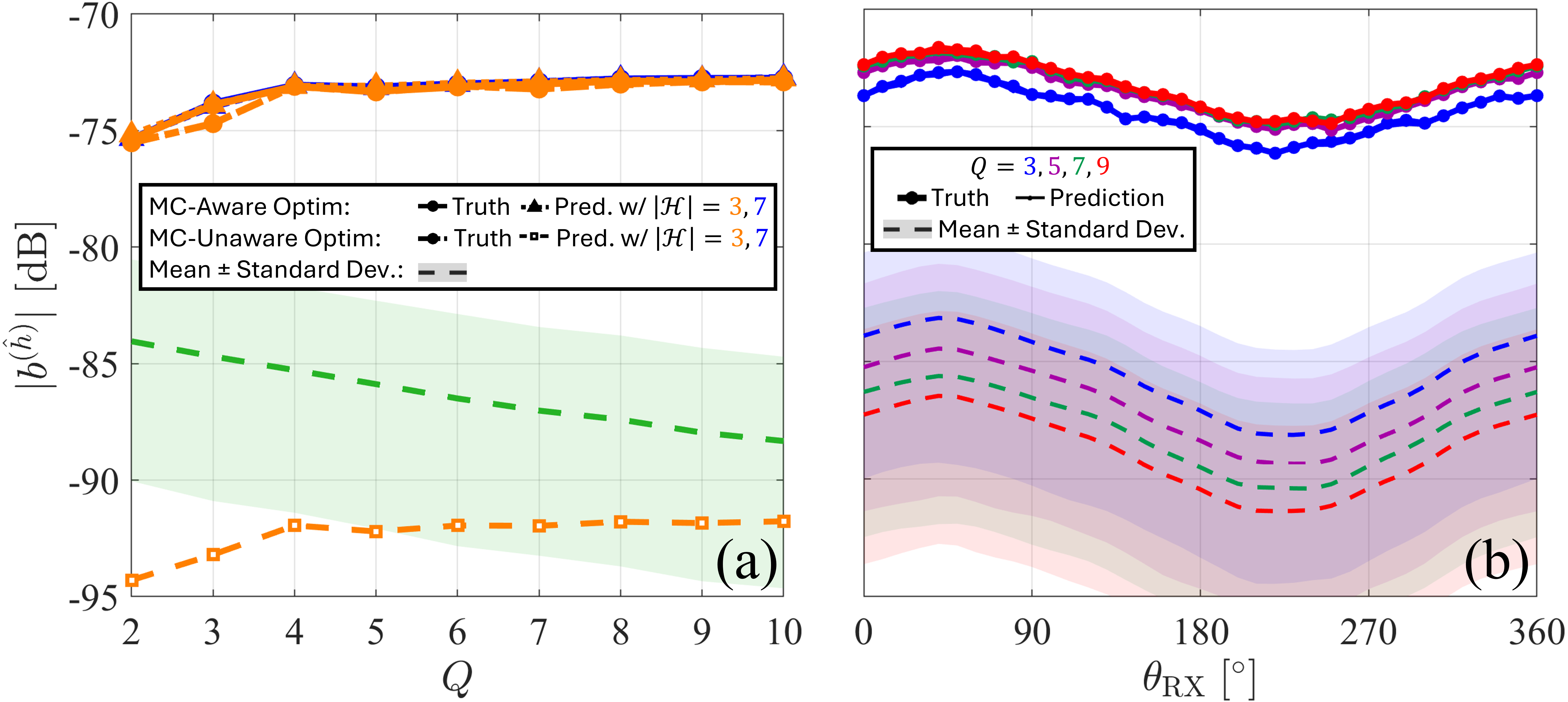}
    \caption{Performance evaluation on \eqref{eq:on_only_opt} as a function of (a) $Q$ (for $\theta_\mathrm{TX}=110^\circ$) and (b) $\theta_\mathrm{RX}$ (for various $Q$, see legend). In (a) we compare optimizations with different $|\mathcal{H}|$ as well as with and without MC awareness. The optimizations in (b) are performed with $|\mathcal{H}|=7$ and MC awareness.}
    \label{Fig4}
\end{figure}

\textit{Performance Evaluation:} We evaluate the TF-RIS performance on \eqref{eq:on_only_opt} as a function of $Q$ and $\theta_\mathrm{RX}$ in Fig.~\ref{Fig4}. Our benchmark is the mean value of $|b^{(\hat{h})}|$ across random TF-RIS modulation patterns. We see in Fig.~\ref{Fig4}(a) that at $Q=2$ our optimization achieves an improvement of roughly 8.7~dB over the benchmark, irrespective of the choice of $|\mathcal{H}|$ and MC awareness during optimization. However, the MC-unaware model in \eqref{eq:HtildeCASC} underestimates the optimized value of $|b^{(\hat{h})}|$ by 18.9~dB. Moreover, the MC-unaware predictions are strictly identical for $|\mathcal{H}|=3$ and $|\mathcal{H}|=7$ because signals in the higher harmonics cannot contribute to the targeted first harmonic via multiple harmonic-scattering events.
For larger values of $Q$, the achieved $|b^{(\hat{h})}|$ only increases marginally to -73.1~dB at $Q=4$ and 
plateaus thereafter. In general, the optimal performance need not improve monotonically with $Q$, except when $Q$ is increased by an integer multiple because in that case  every pattern feasible for the smaller $Q$ is also feasible for the larger $Q$. 
Meanwhile, the benchmark is seen to steadily decrease with $Q$ from -84.0~dB at $Q=2$ to -88.3~dB at $Q=10$. 
Fig.~\ref{Fig4}(b) corroborates the generality for all $\theta_\mathrm{RX}$ of the marginal benefit of larger $Q$, the decline of the benchmark for larger $Q$, and the accuracy of TF-MNT predictions with low $|\mathcal{H}|$. 
We hypothesize that this performance insensitivity to model inaccuracies results from a combination of the binary-tunability constraint, the low value of $N_\mathrm{S}$ in our small-scale TF-RIS, and the optimization objective in \eqref{eq:on_only_opt}. Indeed, a similar insensitivity to model inaccuracies was already observed for conventional small-scale quantized RISs in multiple experiments such as~\cite{del2025experimental}. 

\begin{figure}
    \centering
    \includegraphics[width=\columnwidth]{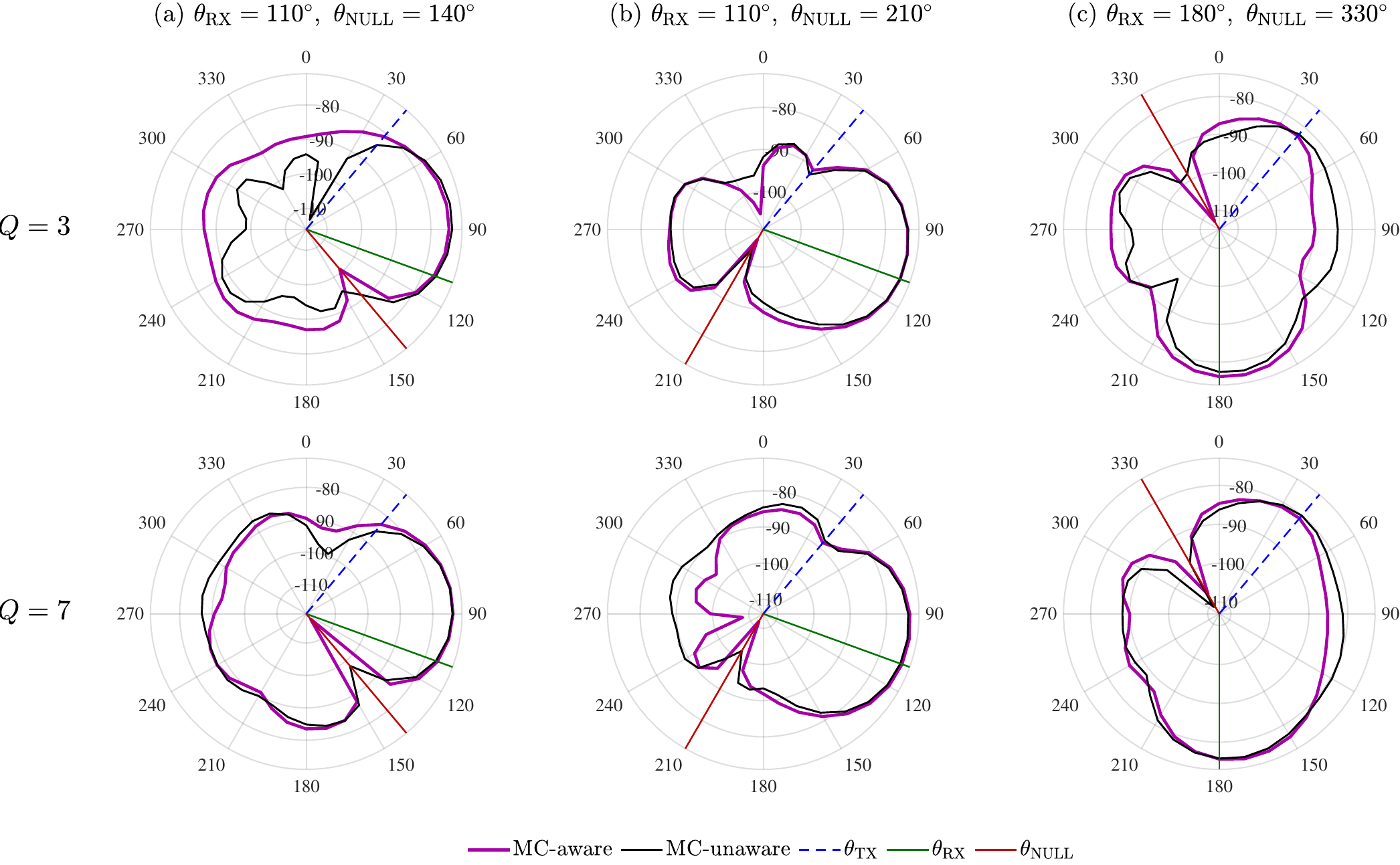}
    \caption{Selected examples of TF-RIS-empowered simultaneous harmonic beam steering and nulling (see \eqref{eq:on_only_opt_beamplusnull}), for $Q=3$ (top row) and $Q=7$ (bottom row). The MC-aware (purple) and MC-unaware (black) optimizations both use $|\mathcal{H}|=11$. The optimized configurations are plotted with $|\mathcal{H}|=51$ and MC awareness. Key metrics are summarized in Table~\ref{tab:polar_gains}. }
    \label{Fig5}
\end{figure}

Next, we examine the TF-RIS performance on \eqref{eq:on_only_opt_beamplusnull} for three selected examples with $Q=3$ and $Q=7$ and $|\mathcal{H}|=11$ in Fig.~\ref{Fig5}. While the desired-channel-gain maximization is largely insensitive to MC awareness and the value of $Q$ in all cases (in line with Fig.~\ref{Fig4}), the undesired-channel-gain minimization displays a notable sensitivity to  MC awareness. MC-unaware optimization appears unable to reliably create the desired null. Moreover, $Q=7$ achieves substantially deeper nulls than $Q=3$ in the first two columns of Fig.~\ref{Fig5}, while the nulls are comparable for $Q=7$ and $Q=3$ in the third column of Fig.~\ref{Fig5}.
These observations corroborate our hypothesis that the nulling aspect of \eqref{eq:on_only_opt_beamplusnull} is more sensitive to MC than the mere channel-gain maximization in \eqref{eq:on_only_opt}, in line with previous experimental examinations of the role of MC awareness~\cite{yven2025end}.

\begin{table}
\caption{Channel gains in Fig.~\ref{Fig5} at $\theta_{\mathrm{RX}}$ and $\theta_{\mathrm{NULL}}$ in dB.}
\label{tab:polar_gains}
\centering
\footnotesize
\setlength{\tabcolsep}{3.2pt}
\renewcommand{\arraystretch}{1.05}
\begin{tabular}{c|c|cc|cc|cc}
\hline
\multirow{2}{*}{$Q$} & \multirow{2}{*}{\makecell[c]{MC-aware\\optimization?}} & \multicolumn{2}{c|}{Fig.~\ref{Fig5}(a)} & \multicolumn{2}{c|}{Fig.~\ref{Fig5}(b)} & \multicolumn{2}{c}{Fig.~\ref{Fig5}(c)} \\
&  & $\theta_{\mathrm{RX}}$ & $\theta_{\mathrm{NULL}}$ & $\theta_{\mathrm{RX}}$ & $\theta_{\mathrm{NULL}}$ & $\theta_{\mathrm{RX}}$ & $\theta_{\mathrm{NULL}}$ \\
\hline
3 & Yes & $-76.7$ & $-101.0$ & $-74.4$ & $-106.2$ & $-76.2$ & $-112.6$ \\
3 & No  & $-76.1$ & $-91.3$ & $-74.4$ & $-103.5$ & $-77.4$ & $-98.1$ \\
\hline
7 & Yes & $-75.9$ & $-116.8$ & $-73.3$ & $-111.6$ & $-75.7$ & $-107.3$ \\
7 & No  & $-76.5$ & $-98.0$ & $-73.7$ & $-102.2$ & $-75.8$ & $-97.9$ \\
\hline
\end{tabular}
\vspace{0.5mm}
\end{table}

\section{Conclusion}

To summarize, we have presented the first MC-aware optimization of a TF-RIS. We formulated the problem based on TF-MNT, we proposed a general and efficient optimization algorithm, and we evaluated the performance based on a numerical full-wave simulation of a practical TF-RIS design. While harmonic channel gain maximization alone was observed to be rather insensitive to model inaccuracies (MC unawareness, low number of retained harmonics) and the time resolution of the periodic modulation, the more demanding optimization goal of simultaneous desired-channel-gain maximization and undesired-channel-gain minimization displayed notable sensitivities.

Looking forward, the application of MC-aware model-based TF-RIS optimization to experiments requires accurate prototype-aware model parameters; to avoid simulation-reality mismatch, generalizing experimental MNT parameter estimation techniques for conventional RISs (e.g.,~\cite{del2025experimental}) to TF-MNT for TF-RISs is an important direction for future work.

\section*{Acknowledgment}
P.d.H. acknowledges stimulating discussions with R.~Jäntti. The authors further acknowledge the MIDAS infrastructure of the Aalto
School of Electrical Engineering.

\bibliographystyle{IEEEtran}
\providecommand{\noopsort}[1]{}\providecommand{\singleletter}[1]{#1}%

\end{document}